# Mechanism of robust circadian oscillation of KaiC phosphorylation *in vitro*

Kohei Eguchi*, Mitsumasa Yoda*, Tomoki P. Terada, and Masaki Sasai

Department of Computational Science and Engineering, Nagoya University, Nagoya 464-8603, Japan

**ABSTRACT**

By incubating the mixture of three cyanobacterial proteins, KaiA, KaiB, and KaiC, with ATP *in vitro*, Kondo and his colleagues reconstituted the robust circadian rhythm of the phosphorylation level of KaiC (*Science*, 308; 414-415 (2005)). This finding indicates that protein-protein interactions and the associated hydrolysis of ATP suffice to generate the circadian rhythm. Several theoretical models have been proposed to explain the rhythm generated in this "protein-only" system, but the clear criterion to discern different possible mechanisms was not known. In this paper, we discuss a model based on the two basic assumptions: The assumption of the allosteric transition of a KaiC hexamer and the assumption of the monomer exchange between KaiC hexamers. The model shows a stable rhythmic oscillation of the phosphorylation level of KaiC, which is robust against changes in concentration of Kai proteins. We show that this robustness gives a clue to distinguish different possible mechanisms. We also discuss the robustness of oscillation against the change in the system size. Behaviors of the system with the cellular or subcellular size should shed light on the role of the protein-protein interactions in *in vivo* circadian oscillation.

*These two authors equally contributed to this work.

**INTRODUCTION**

To resolve the mechanism of circadian rhythms, cyanobacteria have been studied as the simplest organisms to exhibit rhythms. In a cyanobacterium *Synechococcus elongatus* PCC 7942, the gene cluster *kaiABC* and their product proteins, KaiA, KaiB, and KaiC, were shown to be essential in generating the rhythm (1), and intense interest has been focused on these Kai proteins (2-19). In the recent study of Kondo and his colleague (20), the circadian oscillation in the phosphorylation level of KaiC has been reconstituted *in vitro* by incubating the mixture of KaiA, KaiB, and KaiC with ATP. This epoch-making work indicates that protein-protein interactions (21) and the associated hydrolysis of ATP (22) suffice to generate the rhythm in the absence of transcriptional or translational processes.

Interactions between Kai proteins *in vitro* have been characterized in detail (4, 19, 21, 23-26): KaiC has the autophosphatase activity, so that KaiC is gradually dephosphorylated when KaiC alone is incubated *in vitro* (19). KaiC is phosphorylated when KaiC is incubated with KaiA (4, 19, 21), whereas KaiB attenuates the activity of KaiA (4, 21). These observations suggest the scenario that KaiC with the low phosphorylation level interacts with KaiA to make the phosphorylation level high, which then allows interaction between KaiC and KaiB to make the phosphorylation level low. The mixture solution of KaiA, KaiB, and KaiC, therefore should have two phases, the phase of phosphorylation with lower affinity of KaiC to KaiB and the phase of dephosphorylation with higher affinity of KaiC to KaiB. Such phosphorylation dependent interactions among Kai proteins (21, 23) and existence of the two phases (23) were confirmed by experiments. However, the precise mechanism of the robust oscillations remains unclear, which calls for *in silico* modeling of the biochemical network of Kai proteins.

Although the two phases of phosphorylation and dephosphorylation are discernible experimentally, the difference of the system in these two phases is unknown. As KaiC forms hexamer in solution (7, 8, 21), van Zon *et al.* (27) and Yoda *et al.* (28) have assumed that a KaiC hexamer switches between two different structural states in the two phases, each of which has the different affinity to KaiA or KaiB. Figure 1 is the simplest reaction scheme based on this assumption of the allosteric transition of the KaiC hexamer. In this reaction scheme, it is obvious that each individual KaiC hexamer switches between the two phases and therefore shows oscillation in its phosphorylation level. However, because the reactions occur stochastically at different timings, the phosphorylation level of independent KaiC hexamers should be desynchronized, even if all the KaiC hexamers are in the same oscillatory phase at the beginning. Such desynchronization should smear out the oscillation in the phosphorylation level of the ensemble of KaiC hexamers, but the observed clear and robust oscillation of the phosphryration level of the ensemble of many KaiC hexamers strongly suggests that there exists some communication among KaiC hexamers, which synchronizes the phosphorylation level of many KaiC hexamers.

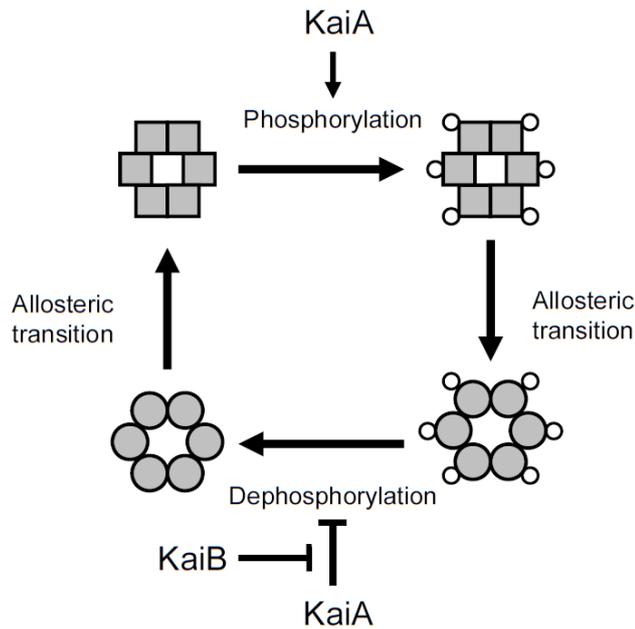

Figure 1
The simplest reaction scheme assuming only the allosteric transition of the KaiC hexamer. The KaiC hexamer undergoes the allosteric transition from one structural state to the other when the phosphorylation level of KaiC is high, and the reverse transition is induced when the phosphorylation level is low. In one structural state, KaiC is phosphorylated by the kinase reaction of KaiA. In the other structural state, KaiC has the higher binding affinity to KaiB, so that KaiB inhibits KaiA from catalyzing phosphorylation. In the latter case, KaiC is dephosphorylated by the autophosphatase reaction of KaiC.

The simplest explanation of this communication would be the complex formation among KaiC hexamers to promote or to suppress phosphorylation or dephosphorylation (29-31). Such interactions should constitute the nonlinear feedback loop, which can stabilize the synchronized oscillation of the phosphorylation level of many of KaiC hexamers. There is, however, no experimental data to support existence of such direct interactions among KaiC hexamers, and hence in this paper, we do not pursue this line.

An important clue can be found in the experimental data that monomers of KaiC are exchanged between KaiC hexamers in the solution of Kai proteins showing the oscillation (21, 24, 26). Ito et al. examined the effects of this monomer shuffling by mixing the oscillatory samples with different phases and found that the synchronization of the oscillation of different samples is mediated by the monomer shuffling during the early dephosphorylation phase (24). We, therefore, may be able to assume that KaiC hexamers can communicate with each other and synchronize their phosphorylation level through the monomer shuffling. In a previous paper, we have conducted a model simulation with a reaction scheme including the monomer shuffling in addition to the allosteric transition of the KaiC hexamer. The result showed that these assumptions are sufficient to sustain the oscillation and to reproduce the experimental data (21) on protein-protein interactions and monomer shuffling semi-quantitatively.

In this paper, robustness of oscillation generated by this mechanism is tested and is compared with experiments. We examine the robustness of oscillation by changing concentrations of Kai proteins and show that our proposed mechanism explains the experimental data on how the oscillation persists in the lower concentrations of Kai proteins. The results also predict the robust oscillation in the higher concentrations of Kai proteins, which should highlight the difference of the present scheme from that based on the other scenario. We also simulate the robustness of the oscillation against the change in the reaction volume down to the size of a sub-cellular domain. Robustness in the small system should shed light on the mechanism of *in vivo* circadian oscillation.

In the following section, we first describe the oscillation observed under the two assumptions of the allosteric transition and the monomer shuffling by using a simplified reaction scheme to illustrate the essential features of thus generated oscillation. Then, to compare the results with experiments quantitatively, the full model with augmented reactions is introduced. With this model, the robustness of the oscillation against the change in the concentration or the reaction volume is discussed. Our assumption of synchronization through the monomer shuffling is compared with other assumptions (26, 27, 32). The last section is devoted to summary and discussion.

**TWO BASIC ASSUMPTIONS: A MINIMAL MODEL**

As discussed in Introduction, two main assumptions in our model are the allosteric transition of KaiC hexamers and the shuffling of KaiC monomers. In order to illustrate the implication of these assumptions, we first discuss a simplified reaction scheme as shown in Figure 2.

We assume that each KaiC hexamer exhibits the allosteric transition between the relaxed (R) and tense (T) states. Noting the fact that the monomer shuffling is more frequent in the dephosphorylation phase than in the phosphorylation phase (21, 24), we may regard that the destabilized KaiC hexamers in the dephosphorylation phase are in the R state, whereas the KaiC hexamers in the phosphorylation phase are in the T state. Each KaiC monomer has two sites to be phosphorylated, Thr432 and Ser431 (5, 6, 23, 25). It was observed that the phosphorylation and dephosphorylation of Thr432 precedes the phosphorylation and dephosphorylation of Ser431 in the oscillation cycle, respectively (23, 25). Though such ordered reactions might play a significant role in generating the rhythm, we here focus, for simplicity, only on Ser431 whose phosphorylation state determines the affinity of KaiC to KaiB (23). Then, the state of each KaiC hexamer is labeled as $R_i$ or $T_i$, where $i = 0\text{-}6$ is the number of monomers with phosphorylated Ser431 in the KaiC hexamer.

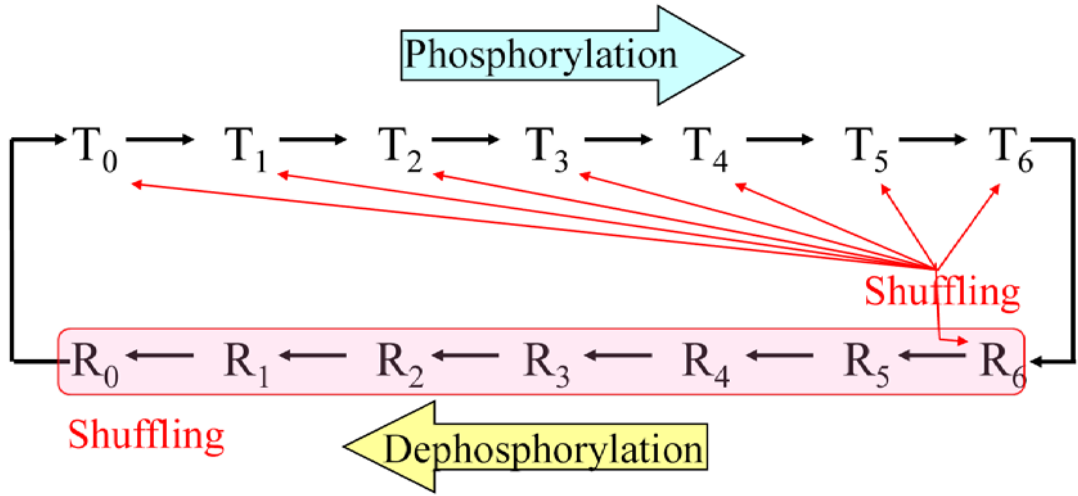

Figure 2
The minimal model assuming both the allosteric transition and the monomer shuffling. Phosphorylation/dephosphorylation and allosteric transitions are indicated by black arrows. The monomer shuffling is allowed between T and $R_6$ (red arrows) and between two Rs (pink area).

KaiA binds to KaiC in the phosphorylation phase, which we identify with the T state. The bound KaiA catalyzes phosphorylation of KaiC (19, 21) as $T_i \to T_{i+1}$. KaiB, on the other hand, binds to KaiC in the dephosphorylation phase, which we identify with the R state, and the bound KaiB inhibits KaiA from catalyzing the phosphorylation of KaiC (4, 21, 23). When KaiB inhibits KaiA from catalyzing the phosphorylation, $i$ decreases due to the autophosphatase activity of the KaiC hexamer (4, 19, 21) as $R_i \to R_{i-1}$. In the present simplified version, the reactions involving KaiA or KaiB are not explicitly taken into account but are effectively represented by the reactions $T_i \to T_{i+1}$ and $R_i \to R_{i-1}$. Since rates of phosphorylation and dephosphorylation are similar to each other (21), we here use the same rate constant for both $T_i \to T_{i+1}$ and $R_i \to R_{i-1}$. If the rate of transition from $T_i$ to $R_i$ is higher for larger $i$ and the rate of the transition from $R_i$ to $T_i$ is higher for smaller $i$, the main population would go through the possible states in a circular manner. In the present simplified version, we regard only $T_6 \to R_6$ and $R_0 \to T_0$ as the effective routes of transition, which is the simplest realization of the circular reactions.

Since the monomer shuffling becomes more frequent in the dephosphorylation phase (21, 24), we assume the shuffling reactions of $R_i + R_j \to R_k + R_l$ with $i + j = k + l$, but the less frequent shuffling of $T_i + T_j \to T_k + T_l$ is not considered here. The monomer shuffling is most frequent at the beginning of the dephosphorylation phase corresponding to the states with large $i$ (24), suggesting that $R_i$ with large $i$ exchanges monomers also with $T_j$s. In the present simplified model, only the shuffling reactions between $R_6$ and $T_j$ are considered. It would be reasonable to assume that KaiC hexamers just after the shuffling tend to be destabilized. In the present model, we simply express this tendency by assuming that KaiC hexamers after the shuffling are in the R states. We thus consider the reactions of $R_6 + T_j \to R_k + R_l$ with $6 + j = k + l$. It should be noted that such shuffling reactions between $R_6$ and $T_j$ have an autocatalytic character since the population of KaiC in the R states increases with the rate which depends nonlinearly on the concentration of KaiC in the $R_6$ states.

By denoting the concentration of $T_i$ and $R_i$ as $[T_i]$ and $[R_i]$, respectively, reactions in the model can be represented as follows;

$$\frac{d[T_0]}{dt} = k_2[R_0] - k_1[T_0] - k_s[T_0][R_6],$$

$$\frac{d[T_i]}{dt} = k_1([T_{i-1}] - [T_i]) - k_s[T_i][R_6], \quad \text{for } i = 1\text{-}5,$$

$$\frac{d[T_6]}{dt} = k_1[R_5] - k_3[T_6] - k_s[T_6][R_6],$$

$$\frac{d[R_6]}{dt} = k_3[T_6] - k_1[R_6] - k_s\sum_{j=0}^{6}[T_j][R_6] - k_s\sum_{j=0}^{5}[R_j][R_6] - 2k_s[R_6]^2 + k_s W_6(\{[R_j]\},\{[T_j]\}),$$

$$\frac{d[R_i]}{dt} = k_1([R_{i+1}] - [R_i]) - k_s\sum_{j=0(j\neq i)}^{6}[R_j][R_i] - 2k_s[R_i]^2 + k_s W_i(\{[R_j]\},\{[T_j]\}),$$
$$\text{for } i = 1\text{-}5,$$

$$\frac{d[R_0]}{dt} = k_1[R_1] - k_2[R_0] - k_s\sum_{j=1}^{6}[R_j][R_0] - 2k_s[R_0]^2 + k_s W_0(\{[R_j]\},\{[T_j]\}), \quad (1)$$

where $k_s W_i(\{[R_j]\},\{[T_j]\})$ represents the rate with which $R_i$ is yielded through shuffling reactions. The explicit forms of the bilinear functions, $W_i(\{[R_j]\},\{[T_j]\})$ with $i$ = 0-6, are given in Appendix. The model consists of 14 variables of $\{[R_i]\}$ and $\{[T_i]\}$ and four rate constants, $k_1$, $k_2$, $k_3$, and $k_S$, where $k_1$ is the rate constant of $T_i \rightarrow T_{i+1}$ and $R_i \rightarrow R_{i-1}$, $k_2$ is the rate constant of $R_0 \rightarrow T_0$, $k_3$ is the rate constant of $T_6 \rightarrow R_6$, and $k_S$ is the rate constant of shuffling reactions. When we scale the unit of time by $1/k_1$ in Eq.1, we can see that the solution of Eq.1 is determined by three independent parameters, $k_2/k_1$, $k_3/k_1$, and $k_S/k_1$.

It should be noted here that the following two conditions are met in the reaction mixture *in vitro*; The monomer shuffling takes place in a relatively short time scale (21), so that the total rate of shuffling, $\sum_i k_S[R_6][T_i]$, are higher than the rate of the allosteric transition, $k_3[T_6]$. On the other hand, as the phosphorylation phase is maintained for a certain duration (23), the rate of the allosteric transition are lower than the rate of phosphorylation. We thus have

$$\sum_i k_S[R_6][T_i] > k_3[T_6],$$
$$k_1[T_i] > k_3[T_6]. \qquad (2)$$

Since the stable oscillation was experimentally observed in a solution containing KaiC monomers of 3.5μM (20, 21, 23), the total concentration of KaiC hexamer is fixed to a constant value of $C_0 = \sum_i ([T_i]+[R_i]) = 3.5\mu M /6 = 0.58\mu M$. Considering $[T_i] \approx [T_6]$ and $C_0 > [R_6]$, Eq.2 should lead to the approximate relations,

$$k_S C_0 > k_3, \quad k_1 > k_3. \qquad (3)$$

In Figure 3, the simulated temporal change of the phosphorylation level,

$$p(t) = \sum_{i=0}^{6} iX_i \bigg/ \left(6\sum_{i=0}^{6} X_i\right), \qquad (4)$$

obtained by numerically integrating Eq.1, is plotted as a function of time, where $X_i =$

$[T_i]+[R_i]$. We should note that even if the individual KaiC hexamers can oscillate their phosphorylation levels, $p(t)$ does not oscillate when the synchronization is not realized among many KaiC hexamers. With the parameters satisfying Eq.3, $p(t)$ in Figure 3 oscillates in a coherent rhythmic manner to show that the simplified scheme as shown in Figure 2 can indeed bring about the synchronization. This synchronization is explained by a series of events occurring in each cycle of the simulated trajectory: First, KaiC accumulates at the $T_6$ state due to the slow transition from T to R. Then, $R_6$ begins to increase by the slow transition of $T_6 \rightarrow R_6$ and when the population of $R_6$ accumulates to some extent, $R_i$ starts to increase rapidly through the autocatalytic reactions of $R_6 + T_j \rightarrow R_k + R_l$. In this way, accumulation of $T_6$ and its collective transition to $R_i$s are the mechanism to synchronize the phosphorylation level of many KaiC hexamers.

To elucidate the importance of the above scenario for the oscillation, the results obtained with the parameters that do not satisfy Eq.3 are also shown in Figure 3. For example, if the shuffling is prohibited with $k_S = 0$, then the mechanism of synchronization does not work and the population of KaiC reaches the stationary distribution which is peaked at around $T_6$. $p(t)$ is kept high due to the accumulation of KaiC hexamers in the $T_6$ state. When $k_3$ is large, on the other hand, the rapid transition from T to R leads to the accumulation of KaiC hexamers in the R states, resulting in a damped oscillation of $p(t)$ which approaches a low constant value. When both conditions of $k_S = 0$ and the large $k_3$ are imposed, $p(t)$ reaches an intermediate value.

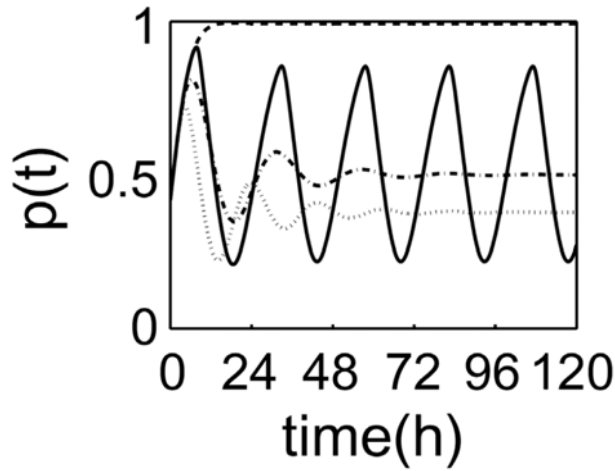

Figure 3

The temporal change of the phosphorylation level $p(t)$ obtained with the minimal model. The solid line is $p(t)$ calculated with the parameterization, $k_2/k_1 = 1$, $k_3/k_1 = 0.001$, and $k_S C_0/k_1 = 10$. $p(t)$ is also simulated by changing one or two of the parameters: the monomer shuffling is prohibited with $k_S = 0$ (dashed line), the allosteric transition from T state to R state is accelerated by increasing $k_3$ to $k_3/k_1 = 1$ (dotted line), both $k_S$ and $k_3$ are changed to $k_S = 0$ and $k_3/k_1 = 1$ (dashed-dotted line). Here, $k_1$ is set to be $1.7 \times 10^{-4} \text{sec}^{-1}$ to fit a period of the real line to the observed value of 22 hours.

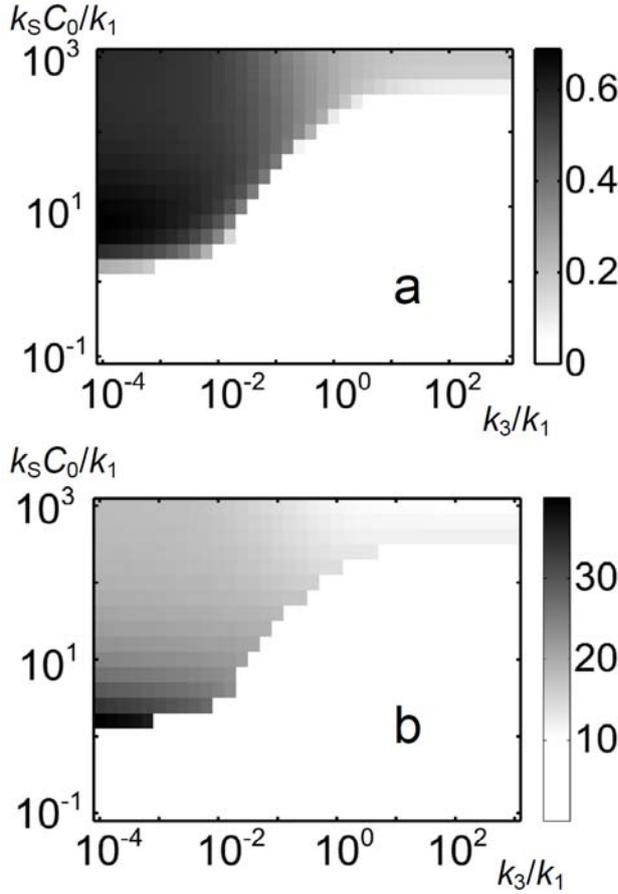

Figure 4
Parameter dependence of solutions of the minimal model. The amplitude (**a**) and the period length (**b**) of the oscillation in $p(t)$ are shown on the plane of $k_3/k_1$ and $k_S C_0/k_1$, where the amplitude is defined as the difference between the maximum and the minimum of oscillating $p(t)$ and the period is shown in unit of hour with $k_1 = 1.7 \times 10^{-4} \text{sec}^{-1}$.

In Figure 4, the amplitude and period of simulated oscillations are shown on the plane of $k_3/k_1$, and $k_S C_0/k_1$. The stable oscillation is found in the parameter region satisfying $k_S C_0 > k_1$ and $k_S C_0 \gg k_3$, which is consistent with the condition of Eq.3. Although the rapid oscillation is also possible in the region of $k_1 < k_3$, the amplitude is small compared to the stochastic noise in the more realistic situation which we will discuss later with the full model. Thus, $k_1 > k_3$ is the region where the stability of oscillation is assured. Qualitative features of Figure 4 are not altered by changing $k_2/k_1$ (data not shown). These results imply that Eq.3 is a necessary condition for the stable oscillation and this should be tested by measuring the rate constants $k_S$ and $k_3$ of KaiC mutants which have lost the rhythmicity and comparing them with those of the wildtype KaiC.

In this section, the mechanism of synchronization of many KaiC hexamers was proposed and illustrated with a minimal model: KaiC hexamers are synchronized by the autocatalytic reactions working through the combined effects of the slow allosteric transition from T to R and the rapid monomer shuffling between T and R. We have also tested the case in which only the shuffling reactions of $R_i + R_j \rightarrow R_k + R_l$ are allowed but the reactions of $R_6 + T_j \rightarrow R_k + R_l$ are prohibited. $p(t)$ does not oscillate in this case (data not shown) as expected from the lack of the autocatalytic mechanism. It has been discussed in Ref.26 that the oscillation can be stabilized when the shuffling reactions of $R_i + R_j \to R_k + R_l$ and $T_i + T_j \to T_k + T_l$ are allowed but the shuffling reactions between $R_i$ and $T_j$ are prohibited, but with such a constraint on shuffling reactions, we did not find a stable oscillation within our model.

**ROBUSTNESS OF OSCILLATION EXAMINED WITH THE FULL MODEL**

To quantitatively assess the robustness of oscillation against the changes in concentration of Kai proteins, the reaction schemes of binding and unbinding among Kai proteins should be explicitly defined. In this section we introduce the full model which incorporates these aspects. In our previous paper (28), we have assumed the Michaelis-Menten kinetics for the binding of KaiA or KaiB to KaiC. With the Michaelis-Menten kinetics, however, the rate of binding is insensitive to changes in concentration of KaiA or KaiB, and such modeling of the predefined insensitivity is not suitable for examining the robustness of oscillation. In the present paper, we do not assume the Michaelis-Menten kinetics but describe the binding of KaiA or KaiB to KaiC as stepwise reactions. With this revision, the robustness of oscillation against changes in concentrations of Kai proteins is examined and its prediction is compared with that based on the other model. We also examine the robustness of oscillation in small systems to compare the simulated results with *in vivo* cyanobacterial oscillation.

**Full model.** KaiA exists as a dimer in solution (11, 14,-16,21), which is denoted here by $A_2$. We assume that a KaiA dimer binds to a KaiC hexamer (16, 21) and a KaiB tetramer, which is denoted by $B_4$, binds to a KaiC hexamer (17,18). We assume that the binding reactions start with the formation of the encounter complexes, $T_i\widetilde{A}_2$, $R_i\widetilde{A}_2$, $T_i\widetilde{B}_4$, or $R_i\widetilde{B}_4$, and then proceed to form the fully bound complexes $T_iA_2$, $R_iA_2$, $T_iB_4$, or $R_iB_4$. KaiA catalyzes the phosphorylation of KaiC (19, 21), and hence the reactions $T_iA_2 \to T_{i+1}A_2$ and $R_iA_2 \to R_{i+1}A_2$ are expected. We also consider the ternary complexes, $T_iA_2\widetilde{B}_4$, $R_iA_2\widetilde{B}_4$, $T_iA_2B_4$ and $R_iA_2B_4$. The other difference from the model used in the previous paper (28) is on the activity of KaiB: In our previous model, KaiB was assumed to work as a phosphatase with the higher dephosphorylation rate of $R_iB_4 \to R_{i-1}B_4$ than that in the autophosphatase reaction of $R_i \to R_{i-1}$. In the present version, we assume that KaiB does not have a strong phosphatase activity but competitively binds to $R_i$ to decrease the concentration of $R_iA_2$, which prevents the kinase reaction of $R_iA_2 \to R_{i+1}A_2$ as was suggested by experiments (4). See *Appendix* for more detailed description of the model used in this paper.

In the full model, dynamical variables are the concentrations of 86 chemical species and 358 chemical reactions among them are taken into account. In order to simplify the model, we assume that each group of similar chemical reactions have the same rate constant, so that the model has 39 rate constants as summarized in *Supporting Table*. These 39 parameters were chosen to satisfy Eq.2 and Eq.3 to maintain the oscillation and were manually calibrated to reproduce the experimental data for kinetics of reactions among Kai proteins *in vitro* (21): The oscillatory period and phase of amount of various complexes of Kai proteins (Figure 2 of Ref.21), the interaction kinetics of KaiA-KaiC and KaiB-KaiC associations (Figure 4 of Ref.21), and the kinetics of shuffling reactions in different conditions (Figure 5 of Ref.21). The ability of the model to semi-quantitatively reproduce these data was discussed in the previous paper (28). In this paper, the main emphasis lies not on the ability of the model to reproduce those data but on the predictions made by thus defined model on the robustness of oscillation. We discuss later in this section that the predicted robustness is intrinsic to the assumptions made in the model and is in sharp contrast to other models based on the different assumptions.

**Stochastic simulation.** When the system exhibits stochastic fluctuations, the fluctuations might destroy the coherent oscillation even when the solution of deterministic equations of law of mass action is oscillatory. To examine the robustness of oscillation, therefore, the stochastic simulation is more suited than the deterministic simulation. We here use the Gillespie algorithm (33) to simulate the stochastic chemical reactions defined in Eqs.5-15 in *Appendix*. In stochastic simulation, chemical reactions are represented by changes in numbers of constituting molecules. We refer to the system consisting of $N_A^0 = 10000$ KaiA dimers, $N_B^0 = 30000$ KaiB dimers, and $N_C^0 = 10000$ KaiC hexamers with the system volume $V^0 = 28.5$ fl as the standard mixture, since the reaction mixture having the corresponding concentrations of 1.2μM KaiA monomers, 3.5μM KaiB monomers, and 3.5μM KaiC monomers has been often used in *in vitro* experiments (20, 21, 23). As shown later in this paper, the oscillation in the small system of $V^0 = 28.5$ fl is stable

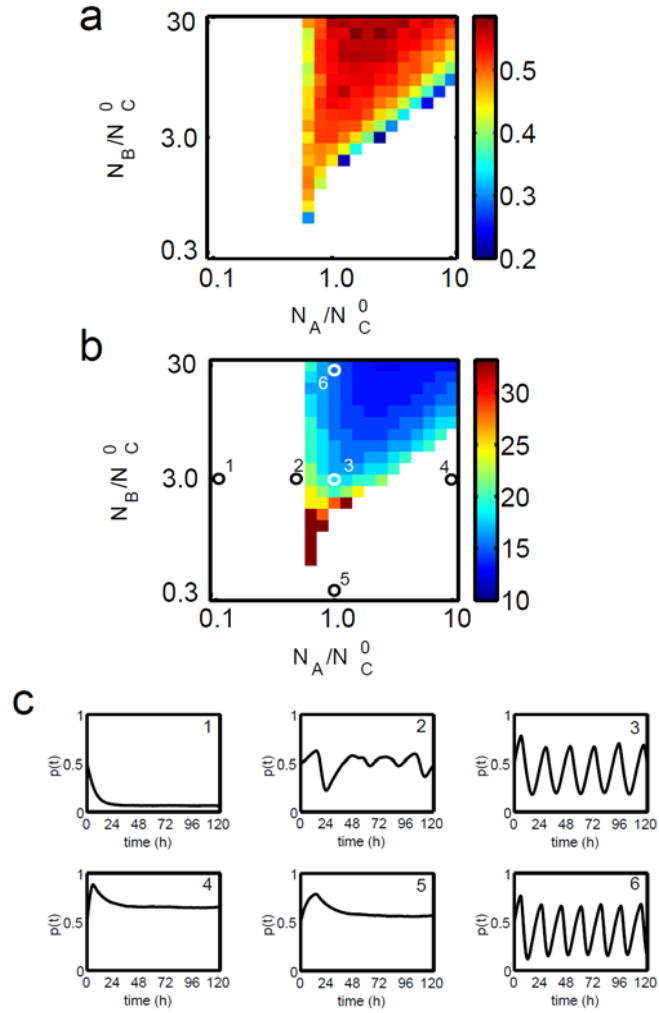

Figure 5

Robustness of the oscillation against changes in concentration of KaiA and KaiB examined with the full model. The amplitude (**a**) and the period length (**b**) of the oscillation in $p(t)$ are shown on the plane of $N_A/N_C^0$ and $N_B/N_C^0$, where the amplitude is defined as the difference between the maximum and the minimum of oscillating $p(t)$ and the period is shown in unit of hour. The numbered panel in **c** shows the trajectory of $p(t)$ calculated with the parameter sets corresponding to the numbered position in **b**: $N_A = 0.126 N_A^0$ and $N_B = N_B^0$ (1), $N_A = 0.5 N_A^0$ and $N_B = N_B^0$ (2), $N_A = N_A^0$ and $N_B = N_B^0$ (3), $N_A = 7.94 N_A^0$ and $N_B = N_B^0$ (4), $N_A = N_A^0$ and $N_B = 0.126 N_B^0$ (5), and $N_A = N_A^0$ and $N_B = 7.94 N_B^0$ (6).

enough to be comparable with the *in vitro* experiments. In the following, we first examine the robustness of oscillation by changing the number of KaiA dimers, $N_A$, the number of KaiB dimers, $N_B$, or the number of KaiC hexamers, $N_C$, by keeping the volume at the standard value of $V = V^0$. We also examine later the robustness of oscillation against change in $V$ by keeping concentrations at the values of the standard mixture as $N_A/V = N_A^0/V^0$, $N_B/V = N_B^0/V^0$, and $N_C/V = N_C^0/V^0$. The total concentration of KaiC hexamer in the phosphorylation level $i$ is calculated as $X_i =$ $[T_i] + [T_i\widetilde{A}_2] + [T_iA_2] + [T_i\widetilde{B}_4] + [T_iB_4] + [T_iA_2\widetilde{B}_4] + [T_iA_2B_4] +$ $[R_i] + [R_i\widetilde{A}_2] + [R_iA_2] + [R_i\widetilde{B}_4] + [R_iB_4] + [R_iA_2\widetilde{B}_4] + [R_iA_2B_4]$ and the phosphorylation level, $p(t)$, is calculated by Eq.4 using thus defined $X_i$.

**Robustness of oscillation against changes in concentration of Kai proteins.** In the simulation of the standard mixture with $N_A = N_A^0$, $N_B = N_B^0$ and $N_C = N_C^0$, $p(t)$ shows a regular coherent oscillation as shown in Fig.5c. To examine the robustness of this oscillation against the changes in $N_A$ and $N_B$ with fixed KaiC concentration of $N_C = N_C^0$ and fixed volume of $V = V^0$, the simulated amplitude and period of oscillation are shown on the plane of $N_A/N_C^0$ and $N_B/N_C^0$ in Figures 5a and 5b. On this plane, there is a distinct region in which the stable coherent oscillation is realized. In the outside of this region, the oscillation dies out in a manner specific for each set of concentrations as shown in Figure 5c. When $N_A/N_C^0$ is small, for example, $p(t)$ reaches a low stationary value due to the insufficient concentration of KaiA for phosphorylating KaiC in the T state. The boundary between the oscillatory region and the region of damped $p(t)$ is at around $N_A/N_C^0 \approx 0.5$-$0.6$, which is consistent with the experimental observation (21). In the region just outside of this boundary, $p(t)$ exhibits chaotic behavior. When $N_A/N_C^0$ is large, on the other hand, the model predicts that $p(t)$ approaches a high stationary value. This is because KaiA exceeds the amount to be inhibited by KaiB, which leads to the sustained highly phosphorylated state of KaiC in the R state. $p(t)$ also reaches a large stationary value when $N_B$ is too small to inhibit the effect of KaiA on KaiC in the R state. When concentration of KaiA is fixed to be $N_A = N_A^0$, the boundary between the oscillatory region and the stationary region is at around $N_B/N_C^0 \approx 1.9$, which is consistent with the experimental observation (21). In contrast, the model predicts that the oscillation is kept stable for large $N_B$ unless the reactions neglected in the model, which would become significant when the concentration of KaiB is very large, could intervene the oscillation. This insensitivity to the increase in the KaiB concentration implies that the stronger inhibition of KaiA activity in the R state due to the larger amount of KaiB does not prevent the system from generating the coherent oscillation.

Thus, we showed that the oscillation is robust over the wide ranges of concentrations: When concentrations of KaiB and KaiC are fixed to be $N_B = N_B^0$ and $N_C = N_C^0$, the oscillation is stable for 0.6< $N_A/N_C$ <3.0, and when concentrations of KaiA and KaiC are fixed to be $N_A = N_A^0$ and $N_C = N_C^0$, the oscillation is stable for 1.9< $N_B/N_C$. The model also predicts that the oscillation is robust when $N_A$ and $N_B$ are increased simultaneously. In this case, the effects of the increase of $N_A$ on KaiC in the R state are compensated by the increase of $N_B$, so that the rate of dephosphorylation of KaiC in the R state is not diminished by the large amount of KaiA.

Robustness was also examined when $N_A$, $N_B$, and $N_C$ are scaled uniformly as $N_A = \alpha N_A^0$, $N_B = \alpha N_B^0$, and $N_C = \alpha N_C^0$. Simulated trajectories of $p(t)$ are shown in Figure 6. Oscillation is stable for large $\alpha$, which is consistent with the observed results (21). When $\alpha$ is small, on the other hand, the rates of phosphorylation and dephosphorylation become low and Eq.1b is not satisfied, with which the stable phosphorylation phase or dephosphorylation phase can not be maintained. The disappearance of the oscillation in the small-$\alpha$ condition is consistent with the experimental data (21).

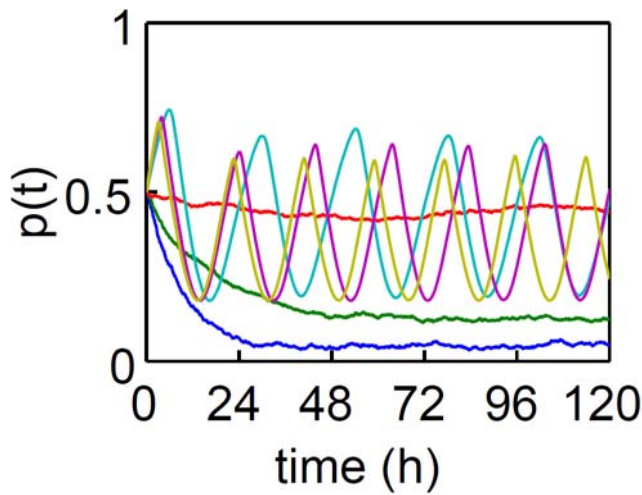

Figure 6

Concentration dependent changes of $p(t)$ calculated with the full model. $p(t)$ for the standard mixture with $\alpha=1$ (light blue), and $p(t)$ for $\alpha = 5.0$ (yellow), $\alpha = 2.5$ (purple), $\alpha = 0.1$ (red), $\alpha = 0.05$ (green), and $\alpha = 0.025$ (blue).

**Comparison with the other scenario.** Robustness of the oscillation demonstrated above gives a useful clue to distinguish the different explanations on the mechanism to generate the rhythm. A simple assumption other than the present assumption of monomer shuffling was put forward by van Zon *et al*. (27) and Takigawa-Imamura and Mochizuki (32). In their assumption, KaiA unbinds more slowly from the less phosphorylated KaiC or KaiA binds more rapidly to the less phosphorylated KaiC, which leads to the differential affinity of KaiA to KaiC. When the KaiA concentration is low, this assumption brings about the shortage of KaiA that can bind to the highly phosphorylated KaiC, which leads to the rapid phosphorylation of the less phosphorylated KaiC. The population of the highly phosphorylated KaiC then accumulates to lead to the synchronization.

We should note that this mechanism of synchronization works only when the amount of free unbound KaiA is small enough to bring about the shortage of KaiA that can bind to the highly phosphorylated KaiC. Indeed, the synchronization is realized in our model by tuning parameters to allow the differential affinity and by prohibiting the monomer shuffling, but the stable oscillation is realized only when $N_A$ is as small as $0.1\,N_A^0$. van Zon *et al*. (27) have introduced the complex $R_iA_4B_4$ in their model ($A_2B_2\overline{C}_i$ in the notation of Ref.27), which should absorb the extra free KaiA without much affecting the kinase/phosphatase kinetics to allow the oscillation under the condition of $N_A \approx N_A^0$. Consistent explanation based on the assumption of differential affinity, therefore, requires existence of the effective absorber of KaiA such as the $R_iA_4B_4$ complex. However, since the extra amount of free KaiA inhibits the synchronization, the oscillation based on differential affinity is fragile when the concentration of KaiA is increased. In Figure 6 of Ref.27, the predicted range of the stable oscillation is $0.4 < N_A/N_C < 1.2$, which is apparently different from the wider range of oscillation $0.6 < N_A/N_C < 3.0$ predicted by the present model. A clearer difference is found when both concentration of KaiA and concentration of KaiB are increased at the same time: In Figure 6 of Ref.27, the predicted range of the stable oscillation does not much depend on the concentration of KaiB and stays as $0.4 < N_A/N_C < 1.2$, which is in sharp contrast to the broadened range of oscillation $0.6 < N_A/N_C < 7.9$ with the increased KaiB concentration of $N_B \approx 2N_B^0$ in the present model.

**Robustness of oscillation against change in the system volume.** Since the cell-cell communication is negligible among cultivated cyanobacterial cells, the observed oscillation of the cyanobacterial clock is the feature of individual cell, which is generated by the intracellular biochemical network (34). Although the circadian rhythm can be reconstituted *in vitro*, the system volume is quite different between *in vitro* and *in vivo* experiments. In the presence of the *in vitro* reconstituted oscillation, one may expect that the same mechanism as in the *in vitro* system works to stabilize the cyanobacterial circadian rhythms *in vivo*. As the system volume becomes smaller down to the cell size, however, the smallness of numbers of molecules in the system should bring about stochastic fluctuations in chemical reactions, which may randomly perturb dynamics to make the oscillation irregular. This raises an important question on whether the reconstituted *in vitro* oscillation can be persistent when the system volume is decreased down to the cellular or sub-cellular size.

We test the persistency of the oscillation by changing the system volume as $\beta V^0$ and the number of Kai proteins $N_A = \beta N_A^0$, $N_B = \beta N_B^0$, and $N_C = \beta N_C^0$ with $\beta < 1$, while keeping the concentrations constant. Persistency of the oscillation in small systems is assessed by measuring the correlation number of cycle, $n_{1/2}$, in the simulated trajectories as follows. Starting from the same initial condition, the time evolution of $p(t)$ in $N = 50$ systems are simulated by $N$ trajectories obtained with different random number seeds. In the $i$th trajectory of simulation, $p(t)$ shows the $n$th peak at time $t = t_i(n)$. Since a first few simulated cycles are transient cycles leaving the features of the initial condition, the data of the first two cycles are not used in taking average, so that the time duration for $n$ cycles is defined by $T_i(n) = t_i(n+2) - t_i(2)$. For $N$ trajectories, we calculate the average time duration for $n$ cycles $T(n) = (1/N)\sum_{i=1}^{N} T_i(n)$ and its variance over different trajectories $\sigma(n)^2 = (1/N)\sum_{i=1}^{N} (T_i(n) - T(n))^2$. When the oscillation is stable, the averaged period length, $L = T(n)/n$, is almost independent of $n$. In Figure 7, the normalized variance, $\sigma(n)^2/L^2$ is plotted as a function of $n$, showing that $\sigma(n)^2$ increases approximately linearly as $n$ increases. The linear dependence of $\sigma(n)^2$ on $n$ indicates that the oscillatory phase shifts stochastically as a random walk in each trajectory. The slope of $\sigma(n)^2/L^2$ is steeper as $\beta$ is smaller, showing that the fluctuation in the oscillatory phase is increased as the system volume is decreased.

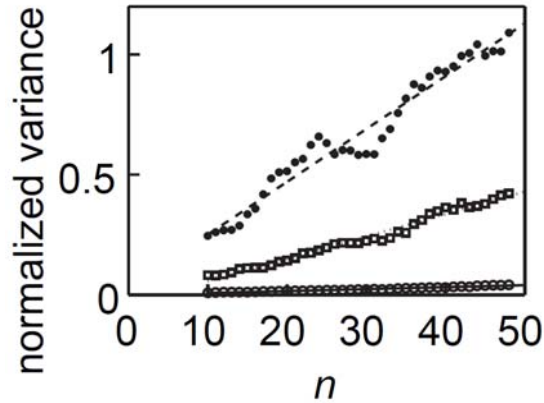

Figure 7

The normalized variance of trajectory-to-trajectory fluctuation of the time length of $n$ cycles, $\sigma(n)^2/L^2$, calculated with the full model is plotted as a function of $n$. $\beta = N_A/N_A^0 = N_B/N_B^0 = N_C/N_C^0 = V/V^0 = 0.398$ (open circles), 0.04 (open squares), and 0.0125 (filled circles) with $N_A^0 = 10000$, $N_B^0 = 30000$, $N_C^0 = 10000$, and $V^0 = 28.5 fl$. The dashed line is a guide for eyes.

As a measure of the timescale for the loss of correlation among different trajectories, we define the correlation number of cycle $n_{1/2}$ by $\sigma(n=n_{1/2})^2/L^2 = (0.5)^2$. In other words, as $n_{1/2}$ days have passed, the trajectory-to-trajectory fluctuation has accumulated to be a half day. In Figure 8, $n_{1/2}$ is plotted as a function of $\beta$. The oscillation is remarkably persistent as $n_{1/2} > 100$ cycles for $\beta > 0.1$. This indicates that the *in vitro* oscillation is quite robust down to the cell size $V = \beta V^0 \approx 1{\sim}2$ fl. Though $N_C$ is as large as $N_C \approx 1000{\sim}2000$ (6000~12000 monomers of KaiC) in a cyanobacterial cell (4), $N_A$ was reported to be as small as $N_A \approx 120{\sim}250$ (240~500 monomers of KaiA) in a cell (4). With this small $N_A$, the reactions have to be confined in a small subcellular region to keep the same concentration as in the standard mixture with $\beta = N_A/N_A^0 = V/V^0 \approx 0.012{\sim}0.025$. For such a small system, $n_{1/2}$ is approximately 10 cycles. In experiment, the correlation time during which the cyanobacterial oscillation is persistent is 166±100 days (35). If the whole bacterial cell volume is used to generate the

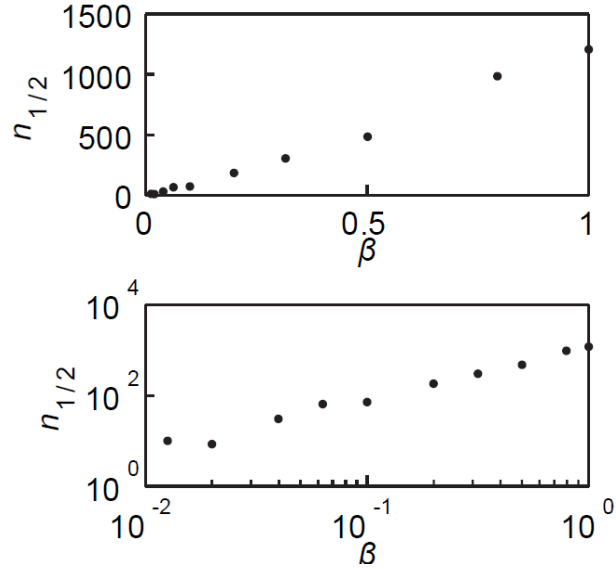

Figure 8

The correlation number of cycles, $n_{1/2}$, calculated with the full model is plotted as a function of $\beta = N_A/N_A^0 = N_B/N_B^0 = N_C/N_C^0 = V/V^0$ with $N_A^0 = 10000$, $N_B^0 = 30000$, $N_C^0 = 10000$, and $V^0 = 28.5$ fl. The upper figure is plotted in the linear scale and the lower figure is in the log-log scale.

oscillation with $N_C \approx 1000\text{~}2000$, $n_{1/2}$ in simulation is comparable with the correlation time observed *in vivo*, and hence the mechanism of the *in vitro* oscillation can sustain the observed single-cell oscillation. If only the subcellular domain is responsible for the oscillation with $N_A \approx N_C \approx 120\text{~}250$ as was suggested in Ref.4, then the persistency of the *in vitro* oscillation is not enough to explain the observed correlation time and the transcriptional-translational feedback mechanism may be needed to complement the persistency. Further theoretical and experimental characterization should help to clarify the relative importance of the robustness provided by three proteins within the observed robustness *in vivo*.

**SUMMARY AND DISCUSSION**

In this paper, we examined the robustness of oscillation generated by the combined effects of the allosteric transition and the monomer shuffling. When the concentration of Kai proteins is decreased, the oscillation generated by this mechanism dies out as was observed in experiments. The model predicted that the oscillation is robust over the wide range of $N_A/N_C$, which should highlight the difference of the proposed mechanism from the mechanism based on the differential affinity. The model also predicted that the oscillation is robust when the system volume is decreased to the cellular size, but when the volume is decreased down to the subcellular size, the other mechanism should be required to explain the observed persistency of the *in vivo* oscillation.

The mechanism presented here may be further extended to conform to the observation that KaiC has two sites, Thr432 and Ser431, to be phosphorylated. When we write the phosphorylated Thr432 and Ser 431 as pT and pS, and nonphosphorylated Thr432 and Ser 431 as T and S, respectively, recent experiments revealed that KaiC undergoes a definite sequence of states as T/S → pT/S → pT/pS → T/pS → T/S in every cycle of oscillation (23, 25).

Rust *et al*. proposed the scenario (25) that KaiC hexamers can be synchronized without relying on the monomer shuffling mechanism by using the sequential process of phosphorylation/dephosphorylation of T and S. Their proposed mechanism is based on the following assumptions (1) The autophosphatase reaction of pT/pS → T/pS is slow and KaiC is accumulated at the pT/pS state because KaiA catalyses the reverse reaction of pT/pS ← T/pS. (2) KaiB does not bind to pT/pS but binds to T/pS. (3) The KaiB-KaiC complex formed in the T/pS state inhibits the activity of KaiA through the formation of KaiABC complex. With these assumptions, KaiC hexamers accumulated in the pT/pS state are collectively transformed to the T/pS state due to the autocatalytic inhibition of the pT/pS ← T/pS reaction through binding of KaiB to KaiC in the T/pS state, and hence the oscillatory phase of KaiC is synchronized. These assumptions, however, seem to be in contradiction to the experimental observation (23) suggesting that (1') The autophosphatase reaction of pT/pS → T/pS is fast and the rate is not affected by the presence or absence of KaiA or KaiB. (2') KaiB binds both to pT/pS and T/pS with the similar binding affinity. (3') KaiB inhibits the reverse reaction of T/pS ← T/S but does not inhibit other reactions significantly. Therefore, at the present stage, we have no consistent model to explain all the reported experimental data and it is especially important to examine whether the mechanism of monomer shuffling proposed in this paper is consistent with the observed data of sequential phosphorylation/dephosphorylation of two sites.

Another important experimental finding is the recent measurement of the ATP hydrolysis reactions catalyzed by KaiC (22). By comparing KaiC mutants, the strong correlation was found between the rate of ATP hydrolysis and the frequency (the inverse of the period length) of the oscillation of the phosphorylation level of KaiC. It is also notable that the rate of ATP hydrolysis is temperature compensated (22). These data strongly suggest that the ATP hydrolysis catalyzed by KaiC is the fundamental reaction which determines the time constant of *in vitro* circadian oscillation. Thus, we now have two basic questions on the mechanism of oscillation of the Kai protein system: One is on the mechanism of the temperature compensated ATP hydrolysis reaction, which determines the fundamental time constant of the KaiC oscillation. A thermodynamic description of the ATP hydrolysis and coupled structural changes should help to clarify the molecular interactions which bring about the temperature compensation. The other is on the mechanism of the synchronization of many KaiC molecules. Although we have discussed only the latter in the present paper, these two problems should be interrelated to each other. Synergetic approaches of both theoretical modeling and experiments should be necessary to find a unified view on these two problems.

**ACKNOWLEDGEMENT**

The authors acknowledge fruitful discussion with Dr. Kondo and his colleagues. The authors also thank Dr. Mihalcescu for discussion on the effects of smallness of the system size. This work was supported by grants from the Ministry of Education, Culture, Sports, Science, and Technology, Japan, and by grants for the 21st century COE program for Frontiers of Computational Science.

**APPENDIX**

**Bilinear functions in the minimal model.** In Eq.1, the bilinear functions $W_i(\{[R_j]\}, \{[T_j]\})$, which determine the rate of yielding $R_i$ through shuffling reactions, are defined as

$$W_0(\{[R_j]\}, \{[T_j]\}) = C_{60}([T_0]+[R_0])[R_6] + (C_{60}[R_1] + C_{50}[R_0])[R_5]$$
$$+ (C_{60}[R_2] + C_{50}[R_1] + C_{40}[R_0])[R_4] + (C_{60}[R_3] + C_{50}[R_2] + C_{40}[R_1] + C_{30}[R_0])[R_3]$$
$$+ (C_{40}[R_2] + C_{30}[R_1] + C_{20}[R_0])[R_2] + (C_{20}[R_1] + [R_0])[R_1] + 2[R_0]^2,$$

$$W_1(\{[R_j]\}, \{[T_j]\}) = (C_{50}[T_1] + C_{61}[T_0] + C_{50}[R_1] + C_{61}[R_0])[R_6]$$
$$+ (C_{50}[R_2] + C_{61}[R_1] + C_{51}[R_0])[R_5] + (C_{50}[R_3] + C_{61}[R_2] + C_{51}[R_1] + C_{41}[R_0])[R_4]$$
$$+ (C_{61}[R_3] + C_{51}[R_2] + C_{41}[R_1] + C_{31}[R_0])[R_3] + (C_{41}[R_2] + C_{31}[R_1] + 2C_{21}[R_0])[R_2]$$
$$+ (2C_{21}[R_1] + [R_0])[R_1],$$

$$W_2(\{[R_j]\}, \{[T_j]\}) = (C_{40}[T_2] + C_{51}[T_1] + C_{62}[T_0] + C_{40}[R_2] + C_{51}[R_1] + C_{62}[R_0])[R_6]$$
$$+ (C_{40}[R_3] + C_{51}[R_2] + C_{62}[R_1] + C_{52}[R_0])[R_5]$$
$$+ (C_{40}[R_4] + C_{51}[R_3] + C_{62}[R_2] + C_{52}[R_1] + 2C_{42}[R_0])[R_4]$$
$$+ (C_{62}[R_3] + C_{52}[R_2] + 2C_{42}[R_1] + C_{31}[R_0])[R_3] + (2C_{42}[R_2] + C_{31}[R_1] + C_{20}[R_0])[R_2]$$
$$+ C_{20}[R_1]^2,$$

$$W_3(\{[R_j]\}, \{[T_j]\}) = (C_{30}[T_3] + C_{41}[T_2] + C_{52}[T_1] + 2C_{63}[T_0])[R_6]$$
$$+ (C_{30}[R_3] + C_{41}[R_2] + C_{52}[R_1] + 2C_{63}[R_0])[R_6]$$
$$+ (C_{30}[R_4] + C_{41}[R_3] + C_{52}[R_2] + 2C_{63}[R_1] + C_{52}[R_0])[R_5]$$
$$+ (C_{41}[R_4] + C_{52}[R_3] + 2C_{63}[R_2] + C_{52}[R_1] + C_{41}[R_0])[R_4]$$
$$+ (2C_{63}[R_3] + C_{52}[R_2] + C_{41}[R_1] + C_{30}[R_0])[R_3] + (C_{41}[R_2] + C_{30}[R_1])[R_2]\},$$

$$W_4(\{[R_j]\}, \{[T_j]\}) = (C_{20}[T_4] + C_{31}[T_3] + 2C_{42}[T_2] + C_{52}[T_1] + C_{62}[T_0])[R_6]$$
$$+ (C_{20}[R_4] + C_{31}[R_3] + 2C_{42}[R_2] + C_{52}[R_1] + C_{62}[R_0])[R_6]$$
$$+ (C_{20}[R_5] + C_{31}[R_4] + 2C_{42}[R_3] + C_{52}[R_2] + C_{62}[R_1] + C_{51}[R_0])[R_5]$$
$$+ (2C_{42}[R_4] + C_{52}[R_3] + C_{62}[R_2] + C_{51}[R_1] + C_{40}[R_0])[R_4]$$
$$+ (C_{62}[R_3] + C_{51}[R_2] + C_{40}[R_1])[R_3] + C_{40}[R_2]^2,$$

$$W_5(\{[R_j]\},\{[T_j]\}) = ([T_5] + 2C_{21}[T_4] + C_{31}[T_3] + C_{41}[T_2] + C_{51}[T_1] + C_{61}[T_0])[R_6]$$

$$+ ([R_5] + 2C_{21}[R_4] + C_{31}[R_3] + C_{41}[R_2] + C_{51}[R_1] + C_{61}[R_0])[R_6]$$

$$+ (2C_{21}[R_5] + C_{31}[R_4] + C_{41}[R_3] + C_{51}[R_2] + C_{61}[R_1] + C_{50}[R_0])[R_5]$$

$$+ (C_{41}[R_4] + C_{51}[R_3] + C_{61}[R_2] + C_{50}[R_1])[R_4] + (C_{61}[R_3] + C_{50}[R_2])[R_3]\},$$

$$W_6(\{[R_j]\},\{[T_j]\}) = (2[T_6] + [T_5] + C_{20}[T_4] + C_{30}[T_3] + C_{40}[T_2] + C_{50}[T_1] + C_{60}[T_0])[R_6]$$

$$+ (2[R_6] + [R_5] + C_{20}[R_4] + C_{30}[R_3] + C_{40}[R_2] + C_{50}[R_1] + C_{60}[R_0])[R_6]$$

$$+ (C_{20}[R_5] + C_{30}[R_4] + C_{40}[R_3] + C_{50}[R_2] + C_{60}[R_1])[R_5]$$

$$+ (C_{40}[R_4] + C_{50}[R_3] + C_{60}[R_2])[R_4] + C_{60}[R_3]^2\},$$

where $C_{ij}$ are factors representing the probability to choose $j$ monomers from $i$ monomers;

$$C_{20} = \frac{{}_2C_0}{{}_2C_0 + {}_2C_1} = \frac{1}{3}, \quad C_{21} = \frac{{}_2C_1}{{}_2C_0 + {}_2C_1} = \frac{2}{3}, \quad C_{30} = \frac{{}_3C_0}{{}_3C_0 + {}_3C_1} = \frac{1}{4}, \quad C_{31} = \frac{{}_3C_1}{{}_3C_0 + {}_3C_1} = \frac{3}{4},$$

$$C_{40} = \frac{{}_4C_0}{{}_4C_0 + {}_4C_1 + {}_4C_2} = \frac{1}{11}, \quad C_{41} = \frac{{}_4C_1}{{}_4C_0 + {}_4C_1 + {}_4C_2} = \frac{4}{11}, \quad C_{42} = \frac{{}_4C_2}{{}_4C_0 + {}_4C_1 + {}_4C_2} = \frac{6}{11},$$

$$C_{50} = \frac{{}_5C_0}{{}_5C_0 + {}_5C_1 + {}_5C_2} = \frac{1}{16}, \quad C_{51} = \frac{{}_5C_1}{{}_5C_0 + {}_5C_1 + {}_5C_2} = \frac{5}{16}, \quad C_{52} = \frac{{}_5C_2}{{}_5C_0 + {}_5C_1 + {}_5C_2} = \frac{10}{16},$$

$$C_{60} = \frac{{}_6C_0}{{}_6C_0 + {}_6C_1 + {}_6C_2 + {}_6C_3} = \frac{1}{42}, \quad C_{61} = \frac{{}_6C_1}{{}_6C_0 + {}_6C_1 + {}_6C_2 + {}_6C_3} = \frac{6}{42},$$

$$C_{62} = \frac{{}_6C_2}{{}_6C_0 + {}_6C_1 + {}_6C_2 + {}_6C_3} = \frac{15}{42}, \quad C_{63} = \frac{{}_6C_3}{{}_6C_0 + {}_6C_1 + {}_6C_2 + {}_6C_3} = \frac{20}{42}.$$

**Full model with the stepwise binding.** We assume that KaiA and KaiC first form an encounter complex of $T_i\widetilde{A}_2$ and then proceed to form a fully bound complex of $T_iA_2$. We also assume that the free energy barrier between $T_iA_2$ and $T_i\widetilde{A}_2$ is higher than that between $T_i\widetilde{A}_2$ and $T_i + A_2$, so that $T_iA_2$ can directly dissociate into $T_i + A_2$ without being trapped into $T_i\widetilde{A}_2$. We have, therefore,

$$T_i + A_2 \rightleftarrows T_i\widetilde{A}_2 \longrightarrow T_iA_2, \quad (5a)$$

$$T_iA_2 \longrightarrow T_i + A_2. \quad (5b)$$

Similarly, for $i = 1\text{-}6$, $R_i$ and KaiA bind in a stepwise way and dissociate in a direct way as

$$R_i + A_2 \rightleftarrows R_i\widetilde{A}_2 \longrightarrow R_iA_2, \quad (6a)$$

$$R_iA_2 \longrightarrow R_i + A_2. \quad (6b)$$

We assume that KaiA does not bind strongly to $R_0$, so that we have

$$R_0 + A_2 \rightleftarrows R_0\widetilde{A}_2. \quad (6c)$$

Eq.6c inherits the assumption used in the model of the previous paper (28). Since phosphorylation of $R_0$ competes with the allosteric transition at $R_0$, the reaction scheme had to be calibrated to control the transition rate by excluding $R_0A_2$ from the model. This calibration was necessary to quantitatively explain the experimental data on the temporal evolution of the shuffling reactions (Figure 8 of Ref.28). We should note that for the simulation of the present paper, the lack of $R_0A_2$ in the model gives little influence to the quantitative results.

    KaiB forms a dimer in solution (11, 21) and KaiC may bind to a tetrameric form of KaiB (17,18). In this model, we assume that dimers and tetramers of KaiB are at equilibrium in solution as

$$2B_2 \rightleftarrows B_4, \quad (7)$$

and KaiC binds to the tetrameric form of KaiB. Since the concentration of the KaiB-KaiC

complex was observed to be high in the dephosphorylation phase (21), we assume the stepwise binding process of $R_i$ and KaiB as

$$R_i + B_4 \rightleftharpoons R_i\tilde{B}_4 \longrightarrow R_iB_4, \qquad (8a)$$

$$R_iB_4 \longrightarrow R_i + B_4. \qquad (8b)$$

In the reaction mixture containing KaiA, KaiB and KaiC, it was observed that the KaiC-KaiA-KaiB tertiary complex is formed both in the phosphorylation and dephosphorylation phases (21), so that we assume

$$T_iA_2 + B_4 \rightleftharpoons T_iA_2\tilde{B}_4 \longrightarrow T_iA_2B_4, \qquad (9a)$$

$$T_iA_2B_4 \longrightarrow T_iA_2 + B_4, \qquad (9b)$$

$$T_iA_2\tilde{B}_4 \longrightarrow T_i + A_2 + B_4, \qquad (9c)$$

$$R_iA_2 + B_4 \rightleftharpoons R_iA_2\tilde{B}_4 \longrightarrow R_iA_2B_4, \qquad (9d)$$

$$R_iA_2B_4 \longrightarrow R_iA_2 + B_4, \qquad (9e)$$

$$R_iA_2\tilde{B}_4 \longrightarrow R_i + A_2 + B_4, \qquad (9f)$$

where Eqs.9c and 9f are based on the assumption that the weak binding of KaiB to the KaiC-KaiA complex does not alter the chemical properties of KaiC-KaiA complex, so that KaiC-KaiA complex dissociates into KaiC and KaiA in the same way as in Eqs.5b and 6b. Due to the autophosphatase activity of KaiC, the phosphorylation level is decreased in the absence of the strongly bound KaiA;

$$T_i \longrightarrow T_{i-1}, \qquad (10a)$$

$$T_i \widetilde{A}_2 \longrightarrow T_{i-1} \widetilde{A}_2, \tag{10b}$$

$$R_i \longrightarrow R_{i-1}, \tag{10c}$$

$$R_i \widetilde{A}_2 \longrightarrow R_{i-1} \widetilde{A}_2, \tag{10d}$$

$$R_i \widetilde{B}_4 \longrightarrow R_{i-1} \widetilde{B}_4, \tag{10e}$$

$$R_i B_4 \longrightarrow R_{i-1} B_4, \tag{10f}$$

where Eqs.10b and 10d are based on the assumption that the weakly bound KaiA in the encounter complex does not catalyze the phosphorylation and KaiC is dephosphorylated in the same way as in Eqs.10a and 10c. In the previous paper (28), we assumed that KaiC is dephosphorylated with a high rate in the KaiC-KaiB complex, but here we assume that KaiB only hinders the binding of KaiA and the rates of the reactions in Eqs.10e and 10f are assumed to be same as that of Eq.10c. Such a role of KaiB was suggested by experiments (4).

When KaiA binds strongly to KaiC in fully bound complex, KaiA catalyzes phosphorylation of KaiC. We assume that the kinase activity of KaiA is not diminished when KaiB additionally binds to the KaiC-KaiA complex through Eqs.9a and 9d;

$$T_i A_2 \longrightarrow T_{i+1} A_2, \tag{11a}$$

$$T_i A_2 \widetilde{B}_4 \longrightarrow T_{i+1} A_2 \widetilde{B}_4, \tag{11b}$$

$$T_i A_2 B_4 \longrightarrow T_{i+1} A_2 B_4, \tag{11c}$$

$$R_i A_2 \longrightarrow R_{i+1} A_2, \tag{11d}$$

$$R_i A_2 \widetilde{B}_4 \longrightarrow R_{i+1} A_2 \widetilde{B}_4, \tag{11e}$$

$$R_i A_2 B_4 \longrightarrow R_{i+1} A_2 B_4. \tag{11f}$$

The allosteric transition from R to T is assumed to be induced also from the weakly bound encounter complexes as

$$R_0 \rightleftarrows T_0, \tag{12a}$$

$$R_0\widetilde{B}_4 \longrightarrow T_0 + B_4, \tag{12b}$$

$$R_0\widetilde{A}_2 \longrightarrow T_0 + A_2. \tag{12c}$$

In Eq.12 we assumed that the transition from $R_0$ to $T_0$ occurs spontaneously with the slower reverse transition from $T_0$ to $R_0$, but the transition from $T_6$ to $R_6$ should not be so fast to maintain the oscillation as explained in the section of *Two Basic Assumptions: A Minimal Model*. We express this slow transition from $T_6$ to $R_6$ by imposing the constraint that the transition is triggered by the binding of KaiB to $T_6$ as

$$T_6 + B_4 \rightleftarrows T_6\widetilde{B}_4 \longrightarrow R_6B_4, \tag{13a}$$

$$T_6\widetilde{A}_2 + B_4 \rightleftarrows T_6\widetilde{A}_2\widetilde{B}_4 \longrightarrow R_6B_4\widetilde{A}_2 \longrightarrow R_6B_4 + A_2, \tag{13b}$$

where Eq.13b is based on the assumption that the weakly bound KaiA does not affect reactions much and the reactions concerning the encounter complex with weakly bound KaiA is similar to those in the KaiA-free state as in Eq.13a. The shuffling reactions are assumed to be similar to those in the simplified version of the section of *Two Basic Assumptions: A Minimal Model*;

$$R_i + R_j \longrightarrow R_k + R_l, \tag{14a}$$

$$R_i\widetilde{B}_4 + R_j \longrightarrow R_k + R_l + B_4, \tag{14b}$$

$$R_i\widetilde{B}_4 + R_j\widetilde{B}_4 \longrightarrow R_k + R_l + 2B_4, \tag{14c}$$

$$R_i\tilde{A}_2 + R_j \longrightarrow R_k + R_l + A_2, \quad (14d)$$

$$R_i\tilde{A}_2 + R_j\tilde{A}_2 \longrightarrow R_k + R_l + 2A_2, \quad (14e)$$

$$R_i\tilde{A}_2 + R_j\tilde{B}_4 \longrightarrow R_k + R_l + A_2 + B_4, \quad (14f)$$

with $i + j = k + l$, and

$$R_6 + T_j \longrightarrow R_k + R_l, \quad (15a)$$

$$R_6\tilde{A}_2 + T_j \longrightarrow R_k + R_l + A_2, \quad (15b)$$

$$R_6\tilde{B}_4 + T_j \longrightarrow R_k + R_l + B_4, \quad (15c)$$

$$R_6 + T_j\tilde{A}_2 \longrightarrow R_k + R_l + A_2, \quad (15d)$$

$$R_6\tilde{A}_2 + T_j\tilde{A}_2 \longrightarrow R_k + R_l + 2A_2, \quad (15e)$$

$$R_6\tilde{B}_4 + T_j\tilde{A}_2 \longrightarrow R_k + R_l + A_2 + B_4, \quad (15f)$$

with $6 + j = k + l$. Here, Eqs.14b-14f and Eqs.15b-15f are based on the assumption that the weak binding of KaiA or KaiB to KaiC does not affect the shuffling reactions. Eqs.5-15 define the reaction schemes of the full model in the present paper. The parameters used in Eqs.5-15 are summarized in *Supplementary Table*.

**REFERENCES**

1. Ishiura, M., S. Kutsuna, S. Aoki, H. Iwasaki, C. R. Andersson, A. Tanabe, S. S. Golden, C. H. Johnson, and T. Kondo. 1998. Expression of a gene cluster kaiABC as a circadian feedback process in cyanobacteria. Science. 281:1519-1523.

2. Iwasaki, H., Y. Taniguchi, M. Ishiura, and T. Kondo. 1999. Physical interactions among circadian clock proteins KaiA, KaiB and KaiC in cyanobacteria. EMBO J. 18:1137-1145.

3. Nishiwaki, T., H. Iwasaki, M. Ishiura, and T. Kondo. 2000. Nucleotide binding and autophosphorylation of the clock protein KaiC as a circadian timing process of cyanobacteria, Proc. Natl. Acad. Sci. USA. 97:495-499.

4. Kitayama, Y., H. Iwasaki, T. Nishiwaki, and T. Kondo. 2003. KaiB functions as an attenuator of KaiC phosphorylation in the cyanobacterial circadian clock system. EMBO J. 22:2127-2134.

5. Nishiwaki, T., Y. Satomi, M. Nakajima, C. Lee, R. Kiyohara, H. Kageyama, Y. Kitayama, M. Temamoto, A. Yamaguchi, A. Hijikata, M. Go, H. Iwasaki, T. Takao, and T. Kondo. 2004. Role of KaiC phosphorylation in the circadian clock system of Synechococcus elongatus PCC 7942. Proc. Natl. Acad. Sci. USA. 101:13927-13932.

6. Xu, Y., T. Mori, R. Pattanayek, S. Pattanayek, M. Egli, and C. H. Johnson. 2004. Identification of key phosphorylation sites in the circadian clock protein KaiC by crystallographic and mutagenetic analyses. Proc. Natl. Acad. Sci. USA. 101:13933-13938.

7. Mori, T., S. V. Saveliev, Y. Xu, W. F. Stafford, M. M. Cox, R. B. Inman, and C. H. Johnson. 2002. Circadian clock protein KaiC forms ATP-dependent hexameric rings and binds DNA. Proc. Natl. Acad. Sci. USA. 99: 17203-17208.

8. Hayashi, F., H. Suzuki, R. Iwase, T. Uzumaki, A. Miyake, J. R. Shen, K. Imada, Y. Furukawa, K. Yonekura, K. Namba, and M. Ishiura. 2003. ATP-induced hexameric ring structure of the cyanobacterial circadian clock protein KaiC. Genes to Cells. 8:287-296.

9. Williams, S. B., I. Vakonakis, S. S. Golden, and A. C. LiWang. 2002. Structure and function from the circadian clock protein KaiA of Synechococcus elongatus: A potential clock input mechanism. Proc. Natl. Acad. Sci. USA. 99:15357-15362.

10. Vakonakis, I., J. Sun, T. Wu, A. Holzenburg, S. S. Golden, and A. C. LiWang. 2004. NMR structure of the KaiC-interacting C-terminal domain of KaiA, a circadian clock protein: Implications for KaiA–KaiC interaction. Proc. Natl. Acad. Sci. USA. 101:1479-1484.

11. Garces, R. G., N. Wu, W. Gillon, and E. F. Pai. 2004. Anabaena circadian clock proteins KaiA and KaiB reveal a potential common binding site to their partner KaiC. EMBO J. 23:1688-1698.

12. Vakonakis, I., and A. C. LiWang. 2004. Structure of the C-terminal domain of the clock protein KaiA in complex with a KaiC-derived peptide: Implications for KaiC regulation. Proc. Natl. Acad. Sci. USA. 101:10925-10930.

13. Xu, Y., T. Mori, and C. H. Johnson. 2003. Cyanobacterial circadian clockwork: roles of KaiA, KaiB and the kaiBC promoter in regulating KaiC. EMBO J. 22: 2117-2126.

14. Ye, S., I. Vakonakis, T. R. Ioerger, A. C. LiWang, and J. C. Sacchettini. 2004. Crystal structure of circadian clock protein KaiA from Synechococcus elongatus. J. Biol. Chem. 279: 20511-20518.

15. Uzumaki, T., M. Fujita, T. Nakatsu, F. Hayashi, H. Shibata, N. Itoh, H. Kato, and M. Ishiura. 2004. Crystal structure of the C-terminal clock-oscillator domain of the cyanobacterial KaiA protein. Nat. Struct. Mol. Biol. 11: 623-631.

16. Pattanayek, R., D. R. Williams, S. Pattanayek, Y. Xu, T. Mori, C. H. Johnson, P. L. Stewart, and M. Egli. 2006. Analysis of KaiA-KaiC protein interactions in the cyano-bacterial circadian clock using hybrid structural methods. EMBO J. 25:2017-2028.

17. Iwase, R., K. Imada, F. Hayashi, T. Uzumaki, M. Morishita, K. Onai, Y. Furukawa, K. Namba, and M. Ishiura. 2005. Functionally important substructures of circadian clock protein KaiB in a unique tetramer complex. J. Biol. Chem. 280:43141-43149.

18. Hitomi, K., T. Oyama, S. Han, A. S. Arvai, and E. D. Getzoff. 2005. Tetrameric architecture of the circadian clock protein KaiB. A novel interface for intermolecular interactions and its impact on the circadian rhythm. J. Biol. Chem. 280:19127-19135.

19. Tomita, J., M. Nakajima, T. Kondo, and H. Iwasaki. 2005. No Transcription－Translation Feedback in Circadian Rhythm of KaiC Phosphorylation. Science. 307:251-254.

20. Nakajima, M., K. Imai, H. Ito, T. Nishiwaki, Y. Murayama, H. Iwasaki, T. Oyama, and T. Kondo. 2005. Reconstitution of Circadian Oscillation of Cyanobacterial KaiC Phosphorylation in Vitro. Science. 308:414-415.

21. Kageyama, H., T. Nishiwaki, M. Nakajima, H. Iwasaki, T. Oyama, and T. Kondo. 2006. Cyanobacterial Circadian Pacemaker: Kai Protein Complex Dynamics in the KaiC Phosphorylation Cycle In Vitro. Molecular Cell. 23:161-171.

22. Terauchi, K., Y. Kitayama, T. Nishiwaki, K. Miwa, Y. Murayama, T. Oyama, and T. Kondo. 2007. ATPase activity of KaiC determines the basic timing for circadian clock of cyanobacteria. Proc. Natl. Acad. Sci. USA. 104:16377-16381.

23. Taeko, N., Y. Satomi, Y. Kitayama, K. Terauchi, R. Kiyohara, T. Takao, and T. Kondo. 2007. A sequential program of dual phosphorylation of KaiC as a basis for circadian rhythm in cyanobacteria. EMBO J. 26: 4029-4037.

24. Hiroshi, I., H. Kageyama, M. Mutsuda, M. Nakajima, T. Oyama, and T. Kondo. 2007. Autonomous synchronization of the circadian KaiC phosphorylation rhythm. Nat. Struct. Mol. Biol. 14:1084-1088.

25. Rust. M. J., J. S. Markson, W. S. Lane, D. S. Fisher, and E. K. O'Shea. 2007. Ordered Phosphorylation Governs Oscillation of a Three-Protein Circadian Clock. Science. 318:809-812.

26. Mori, T., D. R. Williams, M. O. Byrne, X. Qin, M. Egli, H. S. Mchaourab, P. L. Stewart, and C. H. Johnson. 2007. Elucidating the Ticking of an In Vitro Circadian Clockwork. PLoS Biol. 5:841-853.

27. van Zon, J. S., D. K. Lubensky, P. R. H. Altena, and P. R. ten Wolde. 2007. An allosteric model of circadian KaiC phosphorylation. Proc. Natl. Acad. Sci. USA.

104:7420-7425.

28. Yoda, M., K. Eguchi, T. P. Terada, and M. Sasai 2007. Monomer-Shuffling and Allosteric Transition in KaiC Circadian Oscillation. PLoS ONE. 2:e408.

29. Mehra, A., C. Hong, M. Shi, J. J. Loros, and J. C. Dunlap, Ruoff P. 2006. Circadian rhythmicity by autocatalysis. PLoS Comput. Biol . 2:e96.

30. Clodong, S., U. Dühring, L. Kronk, A. Wilde, I. Axmann, H. Herzel, and M. Kollmann. 2007. Functioning and robustness of a bacterial circadian clock. Mol. Syst. Biol. 3:1-9.

31. Emberly, E., and N. S. Wingreen. 2006. Hourglass Model for a Protein-Based Circadian Oscillator. Phys. Rev. Lett. 96:038303.

32. Takigawa-Imamura, H., and A. Mochizuki. Predicting Regulation of the Phosphorylation Cycle of KaiC Clock Protein Using Mathematical Analysis. J. Biol. Rhythms. 21:405-416.

33. Gillespie, D.T. 1977. Exact stochastic simulation of coupled chemical reactions. J. Phys. Chem. 81:2340-2361.

34. Amdaoud, M., M. Vallade, C. W. Schaber, I. Mihalcescu. 2007. Cyanobacterial clock, a stable phase oscillator with negligible intercellular coupling. Proc. Natl. Acad. Sci.USA. 104:7051-7056.

35. Mihalcescu, I., W. Hsing, S. Leibler. Resilient circadian oscillator revealed in individual cyanobacteria. Nature. 2004. 430:81-5.

# Supplementary Table

| Reaction type | Reaction scheme | Rate constants | Values as used in the simulation | Values in standard units |
|---|---|---|---|---|
| Phosphorylation | $T_iA_2 \rightarrow T_{i+1}A_2$ <br> $T_iA_2\tilde{B}_4 \rightarrow T_{i+1}A_2\tilde{B}_4$ <br> $T_iA_2B_4 \rightarrow T_{i+1}A_2B_4$ | $k_p^T$ | $1.4 \times 10^{-2}\ [\Delta t^{-1}]$ | $4.1 \times 10^{-4}\ [\sec^{-1}]$ |
| | $R_iA_2 \rightarrow R_{i+1}A_2$ <br> $R_iA_2\tilde{B}_4 \rightarrow R_{i+1}A_2\tilde{B}_4$ <br> $R_iA_2B_4 \rightarrow R_{i+1}A_2B_4$ | $k_p^R$ | $6.0 \times 10^{-3}\ [\Delta t^{-1}]$ | $1.75 \times 10^{-4}\ [\sec^{-1}]$ |
| Dephosphorylation | $T_i \rightarrow T_{i-1}$ <br> $T_i\tilde{A}_2 \rightarrow T_{i-1}\tilde{A}_2$ | $k_{dp}^T$ | $4.0 \times 10^{-3}\ [\Delta t^{-1}]$ | $1.2 \times 10^{-4}\ [\sec^{-1}]$ |
| | $R_i \rightarrow R_{i-1}$ <br> $R_i\tilde{A}_2 \rightarrow R_{i-1}\tilde{A}_2$ <br> $R_i\tilde{B}_4 \rightarrow R_{i-1}\tilde{B}_4$ <br> $R_iB_4 \rightarrow R_{i-1}B_4$ | $k_{dp}^R$ | $1.1 \times 10^{-2}\ [\Delta t^{-1}]$ | $3.2 \times 10^{-4}\ [\sec^{-1}]$ |
| Association | $T_i + A_2 \rightarrow T_i\tilde{A}_2$ | $k_b^{T\text{-}\tilde{A}}$ | $2.5 \times 10^{-3}\ [\Delta t^{-1}]$ | $1.2 \times 10^6\ [M^{-1}\sec^{-1}]$ |
| | $T_i\tilde{A}_2 \rightarrow T_iA_2$ | $k_b^{T\tilde{A}\text{-}A}$ | $5.0\ [\Delta t^{-1}]$ | $1.5 \times 10^{-1}\ [\sec^{-1}]$ |
| | $T_iA_2 + B_4 \rightarrow T_iA_2\tilde{B}_4$ | $k_b^{TA\text{-}\tilde{B}}$ | $1.0 \times 10^{-1}\ [\Delta t^{-1}]$ | $5.0 \times 10^7\ [M^{-1}\sec^{-1}]$ |
| | $T_iA_2\tilde{B}_4 \rightarrow T_iA_2B_4$ | $k_b^{TA\tilde{B}\text{-}B}$ | $2.5 \times 10^{-1}\ [\Delta t^{-1}]$ | $7.3 \times 10^{-3}\ [\sec^{-1}]$ |
| | $R_i + A_2 \rightarrow R_i\tilde{A}_2$ | $k_b^{R\text{-}\tilde{A}}$ | $2.5 \times 10^{-2}\ [\Delta t^{-1}]$ | $1.2 \times 10^7\ [M^{-1}\sec^{-1}]$ |
| | $R_i\tilde{A}_2 \rightarrow R_iA_2$ | $k_b^{R\tilde{A}\text{-}A}$ | $2.5\ [\Delta t^{-1}]$ | $7.3 \times 10^{-2}\ [\sec^{-1}]$ |
| | $R_iA + B_4 \rightarrow R_iA\tilde{B}_4$ | $k_b^{RA\text{-}\tilde{B}}$ | $1.0 \times 10^{-2}\ [\Delta t^{-1}]$ | $5.0 \times 10^6\ [M^{-1}\sec^{-1}]$ |
| | $R_iA_2\tilde{B}_4 \rightarrow R_iA_2B_4$ | $k_b^{RA\tilde{B}\text{-}B}$ | $2.0 \times 10^{-1}\ [\Delta t^{-1}]$ | $5.8 \times 10^{-3}\ [\sec^{-1}]$ |

| Reaction type | Reaction scheme | Rate constants | Values as used in the simulation | Values in standard units |
|---|---|---|---|---|
| Association | $R_i + B_4 \to R_i\widetilde{B}_4$ | $k_b^{R\text{-}\widetilde{B}}$ | $1.0\times10^{-1}\ [\Delta t^{-1}]$ | $5.0\times10^{7}\ [M^{-1}\sec^{-1}]$ |
| | $R_i\widetilde{B}_4 \to R_iB_4$ | $k_b^{R\widetilde{B}\text{-}B}$ | $8.0\times10^{-1}\ [\Delta t^{-1}]$ | $2.3\times10^{-2}\ [\sec^{-1}]$ |
| | $T_6 + B_4 \to T_6\widetilde{B}_4$ | $k_b^{T_6\text{-}\widetilde{B}}$ | $5.0\times10^{-3}\ [\Delta t^{-1}]$ | $2.5\times10^{6}\ [M^{-1}\sec^{-1}]$ |
| | $T_6\widetilde{A}_2 + B_4 \to T_6\widetilde{A}_2\widetilde{B}_4$ | $k_b^{T_6\widetilde{A}\text{-}\widetilde{B}}$ | $5.0\times10^{-3}\ [\Delta t^{-1}]$ | $2.5\times10^{6}\ [M^{-1}\sec^{-1}]$ |
| | $2B_2 \to B_4$ | $k_b^{2B_2\text{-}B_4}$ | $2.0\times10^{-5}\ [\Delta t^{-1}]$ | $1.0\times10^{4}\ [M^{-1}\sec^{-1}]$ |
| Dissociation | $T_i\widetilde{A}_2 \to T_i + A_2$ | $k_d^{T\widetilde{A}}$ | $3.0\ [\Delta t^{-1}]$ | $8.7\times10^{-2}\ [\sec^{-1}]$ |
| | $T_iA_2 \to T_i + A_2$ $T_iA_2\widetilde{B}_4 \to T_i + A_2 + B_4$ | $k_d^{TA}$ | $5.0\ [\Delta t^{-1}]$ | $1.5\times10^{-1}\ [\sec^{-1}]$ |
| | $T_iA_2\widetilde{B}_4 \to T_iA_2 + B_4$ | $k_d^{T\widetilde{B}}$ | $1.0\times10^{-1}\ [\Delta t^{-1}]$ | $2.9\times10^{-3}\ [\sec^{-1}]$ |
| | $T_iA_2B_4 \to T_iA_2 + B_4$ | $k_d^{TB}$ | $4.0\ [\Delta t^{-1}]$ | $1.2\times10^{-1}\ [\sec^{-1}]$ |
| | $R_i\widetilde{A}_2 \to R_i + A_2$ | $k_d^{R\widetilde{A}}$ | $2.0\times10^{-1}\ [\Delta t^{-1}]$ | $5.8\times10^{-3}\ [\sec^{-1}]$ |
| | $R_iA_2 \to R_i + A_2$ $R_iA_2\widetilde{B}_4 \to R_i + A_2 + B_4$ | $k_d^{RA}$ | $9.0\times10^{-1}\ [\Delta t^{-1}]$ | $2.6\times10^{-2}\ [\sec^{-1}]$ |
| | $R_iA_2\widetilde{B}_4 \to R_iA_2 + B_4$ | $k_d^{R\widetilde{B}}$ | $1.0\times10^{-1}\ [\Delta t^{-1}]$ | $2.9\times10^{-3}\ [\sec^{-1}]$ |
| | $R_iA_2B_4 \to R_iA_2 + B_4$ | $k_d^{RB}$ | $2.5\ [\Delta t^{-1}]$ | $7.3\times10^{-2}\ [\sec^{-1}]$ |
| | $R_i\widetilde{B}_4 \to R_i + B_4$ | $k_d^{R\widetilde{B}}$ | $4.0\times10^{-1}\ [\Delta t^{-1}]$ | $1.2\times10^{-2}\ [\sec^{-1}]$ |
| | $R_iB_4 \to R_i + B_4$ | $k_d^{RB}$ | $1.8\ [\Delta t^{-1}]$ | $5.2\times10^{-2}\ [\sec^{-1}]$ |

| Reaction type | Reaction scheme | Rate constants | Values as used in the simulation | Values in standard units |
|---|---|---|---|---|
| Dissociation | $T_6\tilde{B}_4 \rightarrow T_6 + B_4$ | $k_d^{T_6\tilde{B}}$ | $1.0\,[\Delta t^{-1}]$ | $2.9\times10^{-2}\,[\text{sec}^{-1}]$ |
| | $T_6\tilde{A}_2\tilde{B}_4 \rightarrow T_6\tilde{A}_2 + B_4$ | $k_d^{T_6\tilde{B}}$ | $1.0\,[\Delta t^{-1}]$ | $2.9\times10^{-2}\,[\text{sec}^{-1}]$ |
| | $R_6B_4\tilde{A}_2 \rightarrow R_6B_4 + A_2$ | $k_d^{R_6\tilde{B}}$ | $1.0\times10^{-1}\,[\Delta t^{-1}]$ | $2.9\times10^{-3}\,[\text{sec}^{-1}]$ |
| | $B_4 \rightarrow 2B_2$ | $k_d^{B_4}$ | $1.0\times10^{-1}\,[\Delta t^{-1}]$ | $2.9\times10^{-3}\,[\text{sec}^{-1}]$ |
| Allosteric transition | $T_6\tilde{B}_4 \rightarrow R_6B_4$ | $k_{T\rightarrow R}^{T_6\tilde{B}}$ | $5.0\times10^{-5}\,[\Delta t^{-1}]$ | $1.5\times10^{-6}\,[\text{sec}^{-1}]$ |
| | $T_6\tilde{A}_2\tilde{B}_4 \rightarrow R_6B_4\tilde{A}_2$ | $k_{T\rightarrow R}^{T_6\tilde{A}\tilde{B}}$ | $5.0\times10^{-5}\,[\Delta t^{-1}]$ | $1.5\times10^{-6}\,[\text{sec}^{-1}]$ |
| | $R_0 \rightarrow T_0$ | $k_{R\rightarrow T}^{R_0}$ | $6.0\times10^{-3}\,[\Delta t^{-1}]$ | $1.75\times10^{-4}\,[\text{sec}^{-1}]$ |
| | $T_0 \rightarrow R_0$ | $k_{T\rightarrow R}^{T_0}$ | $2.0\times10^{-3}\,[\Delta t^{-1}]$ | $5.8\times10^{-5}\,[\text{sec}^{-1}]$ |
| | $R_0\tilde{A}_2 \rightarrow T_0 + A_2$ | $k_{R\rightarrow T}^{R_0\tilde{A}}$ | $6.0\times10^{-3}\,[\Delta t^{-1}]$ | $1.75\times10^{-4}\,[\text{sec}^{-1}]$ |
| | $R_0\tilde{B}_4 \rightarrow T_0 + B_4$ | $k_{R\rightarrow T}^{R_0\tilde{B}}$ | $6.0\times10^{-3}\,[\Delta t^{-1}]$ | $1.75\times10^{-4}\,[\text{sec}^{-1}]$ |
| Shuffling | $R_i + R_j \rightarrow R_k + R_l$ | $k_s^{RR}$ | $1.0\times10^{-4}\,[\Delta t^{-1}]$ | $5.0\times10^{4}\,[\text{M}^{-1}\text{sec}^{-1}]$ |
| | $R_6 + T_i \rightarrow R_k + R_l$ | $k_s^{RT}$ | $2.0\times10^{-5}\,[\Delta t^{-1}]$ | $1.0\times10^{4}\,[\text{M}^{-1}\text{sec}^{-1}]$ |